\documentstyle[12pt,graphicx]{article}
\title{ 2-loop Quantum Yang-Mills Condensate as Dark Energy  }
\author{T.Y. Xia  and  Y. Zhang\cite{email} \\
        Astrophysics Center \\
        University of Science and Technology of China \\
        Hefei, Anhui, China }
 \date{}

\begin{document}
\maketitle
\baselineskip=21truept

\newcommand{\be}{\begin{equation}}
\newcommand{\ee}{\end{equation}}

\sf
\begin{center}
\Large  Abstract
\end{center}
\begin{quote}
 {\large
In seeking a model solving the coincidence problem,
the effective Yang-Mills condensate (YMC) is an alternative
candidate for dark energy.
A study is made for the model
up to the 2-loop order of quantum corrections.
It is found that, like in the 1-loop model,
for generic initial conditions during the radiation era,
there is always a desired tracking solution,
yielding the current status $\Omega_\Lambda \simeq 0.73$
and $\Omega_m \simeq 0.27$.
As the time $t\rightarrow \infty$
the dynamics is a stable attractor.
Thus the model naturally solves the coincidence problem
of dark energy.
Moreover, if YMC decays into matter,
its equation of state (EoS)
crosses -1 and takes $w\sim -1.1$,
as indicated by the recent observations. }

\end{quote}



e-mail: yzh@ustc.edu.cn

\newpage
%\twocolumn
\baselineskip=21truept

\begin{center}
{\em\Large 1. Introduction}
\end{center}

The energy content of the present Universe is
such that $\Omega _{\Lambda }\simeq0.73$ and
$\Omega _m\simeq0.27$, as indicated by the observational
results from SN Ia \cite{perlmutter}
and  CMB anisotropies \cite{spergel},
and from studies of the large scale structure as well \cite{bahcall}.
The origin of the cosmic dark energy is
now a big challenge to astronomy and physics.
There have been several types of models proposed
to interpret the dark energy.
The simplest one is the cosmological constant $\Lambda$.
Since the matter density evolves as  $\rho_m \propto a(t)^{-3}$
and $\rho_\Lambda\sim$ constant
throughout the history of the Universe,
the initial value of  $\rho_\Lambda/\rho_m$ has to be
chosen with great precision to achieve $\rho_m \sim 0.37\rho_\Lambda$ today.
This  is  the  coincidence problem \cite{coin}.
Moreover, if $\Lambda $ were interpreted to arise
from the vacuum fluctuations of quantum fields \cite{feynman},
then there is a ``fine-tuning" difficulty,
i.e., why at present the vacuum energy of a scale $10^{-3}$ ev
is so tiny compared to the typical scales in particle physics.
Another model is the Steady State Universe \cite{bondi},
in which the accelerating expansion
is driven by some $C$-field with negative energy,
difficult to accept as a physical field.
The effective gravity \cite{raval} is a model
outside the scope of General Relativity,
which has a task to to explain all observed features of gravity.
The Born-Infeld quantum condensate \cite{Elizalde} is a model,
which uses gauge fields as a candidate for the dark energy.
One interesting speculations is that
the dark energy may result from some  dynamic field
with a tracker behavior,
i.e., the field is subdominant during early stages of expansion,
later becomes dominant as the dark energy.
Among this scalar kind of models are the quintessence
 \cite{quint},
k-essence \cite{k}, phantom \cite{phantom}, quintom\cite{quintom}, etc.
For a dynamic field model to solve the coincidence problem,
it is required not to spoil the standard Big Bang cosmology,
i.e.,
the Big Bang nucleosynthesis and the recombination
must occur as usual,
and the matter era must be long enough for structure formation.
Moreover, the solution needs to be a stable tracker,
insensitive to the initial conditions.
But so far the scalar models have difficulty
to fulfil these criteria  \cite{tsujikawa}.

In our previous work \cite{zhang0} \cite{zhang06} \cite{zhao06}
a dynamic model is proposed,
in which the effective YMC,
up to 1-loop quantum corrections, serves as the dark energy.
For quite generic initial conditions
the model always has the desired tracking  behavior,
 naturally solves the coincidence problem,
in the sense that it satisfies those criteria mentioned above.
When YMC is coupled with matter,
EoS of YMC crosses -1 and takes $w\sim -1.1$,
as indicated by the fittings of
recent preliminary observational data
\cite{Astier} \cite{vasey} \cite{riess06} \cite{alam}.
One may ask, does the model still work for more quantum corrections?
As will be seen in this paper,
the YMC dark energy model including the 2-loop quantum corrections
has a tracking solution that solves the coincidence problem,
like the 1-loop model.
Through out the paper  we use the units with $c=\hbar=1$.

\begin{center}
{\em\Large 2. The Effective Lagrangian of 2-Loop YM Condensate}
\end{center}

The quantum effective YM condensate is described by
the following  Lagrangian up to 2-loops
\cite{pagels,adler}
\be \label{Lagrangian}
L_{eff}=\frac{b}{2}F
  \left[ \ln|\frac{F}{e\kappa^2}| +
  \eta \ln|\ln|\frac{F}{e\kappa^2}|+\delta| \right],
\ee
where, for the gauge group $SU(N)$ without fermions,
$b=\frac{11N}{3(4\pi)^2}$ representing the 1-loop part,
and $\eta \equiv \frac{2b_1}{b^2} \simeq 0.84$ with
$b_1=\frac{17N^2}{3(4\pi)^4}$
representing the 2-loop contribution,
the parameter $\kappa$ is the renormalization
scale with dimension of squared mass,
and $F\equiv -\frac{1}{2} F^a{}_{\mu\nu} F^{a}{}^{\mu\nu}= E^2-B^2$
plays the role of the order parameter of the YM condensate.
We consider only the case of ``electric'' condensate with $F=E^2$.
Here  the dimensionless constant  $\delta$ is a parameter
representing higher order corrections.
Eq.(\ref{Lagrangian}) is equivalent to having
the running coupling constant \cite{caswell}
\be \label{g}
g^2(F) =   \frac{1}{b   \tau}
    - \frac{2b_1}{b^3} \frac{ \ln|\tau |}{\tau^2}
    +O(\frac{1}{\tau^2}),
\ee
with $\tau\equiv  \ln|F/e\kappa^2|$ for $L_{eff}=F/2g^2(F)$.
$L_{eff}$ in Eq.(\ref{Lagrangian}) maintains the gauge invariance
and the Lorentz invariance,
and yields the correct trace anomaly \cite{Collins} \cite{pagels}.
Eq.(\ref{Lagrangian}) is reminiscent  of
of the Coleman-Weinberg effective potential
of scalar fields \cite{coleman}.
At strong fields, the corrections to the brackets in
Eq.(\ref{Lagrangian}) are of order unity and,
to the extent that they slowly vary,
can be either absorbed into the scale mass $\kappa^{1/2}$,
or represented by the parameter $\delta$.
As is explained in Ref.\cite{adler},
renormalization group estimates
are formally valid whenever the running coupling $g^2(F)$
is small in magnitude,
(i.e.,  $F/\kappa^2\gg 1$, giving $g^2(F)$  positive,
and $F/\kappa^2\ll 1$, giving $g^2(F)$  negative).
This suggests that there is
a second weak-field asymptotically free regime,
where Eqs.(\ref{Lagrangian}) and (\ref{g})
give the leading behavior,
and where $g^2(F)$ is negative,
a feature of the effective YM condensate \cite{zhang0} \cite{adler2}.
By the way,  the YMC  introduced here is
not the gluon fields in QCD,
nor the gauge boson fields in the electro-weak unification.

The energy density and the pressure, $\rho_y$ and $p_y$,
 of YMC  are given by
\be \label{energy}
\rho_y=\frac{b}{2} F \left\{ \tau +2
+ \eta \left( \ln| \tau+\delta|
  +\frac{2}{ \tau +\delta} \right)  \right\},
\ee
\be \label{pressure}
p_y=   \frac{b}{6} F  \left\{  \tau-2
 + \eta \left(\ln|\tau+\delta|
 -\frac{2}{ \tau+\delta}  \right) \right\}.
\ee
So the form of the stress tensor $T_{\mu\nu}$
of YM fields is consistent with homogeneity and isotropy of the Universe.
The EoS for the YMC is given by
\be \label{w}
w=\frac{p_y}{\rho_y} \, .
\ee
When one sets $\eta=0$ in the above expressions,
the 1-loop model is recovered \cite{zhang06}.
For comparison,
we plot $\rho_y$, $p_y$, and $w$ in  Fig.\ref{fig1}
for both the 2-loop and the 1-loop order, respectively,
where the variable is $y\equiv\tau+1=\ln |F/\kappa^2|$.
We see that $\rho_y$, $p_y$, and $w$ in
the 2-loop order have similar shapes to, and higher magnitude than,
those in the 1-loop one, respectively.
At high energies $y\rightarrow \infty$, $\rho_y$ and $p_y$ are positive,
and the EoS of YMC approaches to that of a radiation, $w\rightarrow 1/3$,
as is expected for an effective theory.
At low energies $y<2.3$, $p_y$ and $w$ become negative,
and at $y<-0.8$ the weak energy
condition \cite{zhang0} \cite{hawking} \cite{parker-zhang}
is violated,
$\rho_y+p_y<0$,  and  $w$ crosses  $-1$.
In particular,
within the range of  $y\simeq (-1.04,\, 54)$  attainable
in our model of cosmic dynamic evolution,
these physical quantities  are all smooth functions.
Although  $g^2$  formally encounters a divergence at $y \sim 0.4$,
nevertheless, the physical quantities, $\rho_y$, $p_y$ and  $w$,
actually involved in our dynamical model are smooth,
a similar situation as in the 1-loop model.
The scale $\kappa$ can be fixed by requiring
$ \rho_y$ in Eq.(\ref{energy})
be equal to the dark energy density  $\sim 0.73\rho_c$,
where $\rho_c$ is the critical density,  yielding
$\kappa^{1/2} \simeq  7.7 \, h_0^{1/2}\times 10^{-3} eV$,
where $h_0$ is the Hubble parameter.
At the moment we do not have an answer to
the question why $\kappa$ is so small,
so the ``fine-tuning" problem is also present in our model.

\begin{center}
{\em\Large 3. The Cosmic Evolution By the YM Condensate}
\end{center}

The spacetime of the universe is described
by  a spatially flat Robertson-Walker metric
\be \label{metric}
 ds^2= d t^2-a^2(t) \delta_{ij} dx^i dx^j ,
 \ee
and is filled with the dark energy,
the matter,  including both  baryons and dark matter,
and the radiation.
In our model,  the dark energy component is represented by the YMC.
To be concordant  with the isotropy of the Universe,
one may take an $SU(2)$ YM field with a highly symmetric configuration
$A^a_0=0$ and $A^a_i=\phi(t) \delta_{ai}$,
where $\phi(t)$ is a scalar function and $a=1,2,3$ \cite{zhang1994}.
The overall cosmic expansion
is determined by the  Friedmann equations
\be \label{friedmann1}
(\frac{\dot{a}}{a})^2=\frac{8 \pi G}{3}(\rho_y+\rho_m+\rho_r),
\ee
where $\rho_m$ and $\rho_r$ are the energy density of
the matter and radiation components respectively.
The dynamical evolutions of the three components are
\be \label{ymeq}
\dot{\rho}_y+3\frac{\dot{a}}{a}(\rho_y+p_y)=-\Gamma \rho_y,
\ee
\be \label{meq}
\dot{\rho}_m+3\frac{\dot{a}}{a}\rho_m=\Gamma \rho_y,
\ee
\be \label{req}
\dot{\rho}_r+3\frac{\dot{a}}{a}(\rho_r+p_r)=0,
\ee
where  $\Gamma$ is the decay rate of the YMC into matter,
a parameter of the model.
Since $\Gamma>0$,
so the term $\Gamma\rho_y$ in Eqs.(\ref{ymeq}) and (\ref{meq})
represents the rate of energy transfer from the YMC component to
the matter one.
Here for simplicity
we do not consider the coupling between the YMC and the radiation.
The sum of Eqs.({\ref{ymeq}}), ({\ref{meq}}), and ({\ref{req}})
guarantees  that the total energy of the three components
is still conserved.
In terms of the variable  $N \equiv\ln a(t)$,
and of the re-scaled functions $y$,
$x \equiv \rho_m/ \frac{1}{2}b \kappa^2$,
and $r\equiv \rho_r/ \frac{1}{2}b \kappa^2$,
the set of equations, (\ref{friedmann1}) through ({\ref{req}}),
take the following form:
\be \label{N}
\dot{N}^2= H^2h^2,
\ee
\be \label{y}
\frac{dy}{d N}=\frac{-4y+ 4\eta (\ln|y-1+\delta|+\frac{1}{y-1+\delta})
-\frac{\gamma}{h}[y+1+\eta(\ln|y-1+\delta|
+\frac{2}{y-1+\delta})]  }{   y+2+\eta (\ln|y-1+\delta|
+\frac{3}{y-1+\delta}-\frac{2}{(y-1+\delta)^2}) },
\ee
\be \label{x}
\frac{dx}{d N}= -3x+\frac{\gamma}{h}
   \left[ y+1+\eta(\ln|y-1+\delta|+\frac{2}{y-1+\delta}) \right],
\ee
\be \label{r}
\frac{dr}{d N}= -4r,
\ee
where  $\gamma \equiv\Gamma/H$,
 $H\equiv\sqrt{4 \pi Gb\kappa^2/3}$,
$h=\sqrt{x+r+e^y[y+1+\eta(\ln|y-1+\delta|+\frac{2}{y-1+\delta})]}$.
Looking at the forms, these equations are much more complicated
than those in the 1-loop  model with $\eta=0$.
Once the parameters  $\gamma$ and $\delta$,
as well as the initial conditions,  are specified,
the solution of this set of equations follows straightforwardly.

As an example,
we take the parameter $\delta=3$ and the decay rate $\gamma=0.5$.
The initial conditions for
the set of  Eqs.(\ref{N}) through (\ref{r})
are chosen at a very high redshift
$z_i= 10^8$ during the radiation era.
To ensure the equality of radiation-matter
occurring  at a redshift $z= 3454$ \cite{spergel},
the initial radiation and matter are taken as
\be \label{xi}
x_i=3.68 \times 10^{22},  \,\,\,\, r_i= 7.25\times  10^{26}.
\ee
To ensure the Big-Bang nucleosynthesis \cite{walker},
the initial YMC fraction should be $\sim 10\% $ or less \cite{zhao06}.
For concreteness we take it to  be
\be \label{yi}
y_i \leq   54,  \,\,\,\,\,  {\rm i.e.},  \,\,\,\,\,
\frac{\rho_{yi}}{\rho_{ri}}\leq     2.28\times 10^{-2}  .
\ee
The solutions for $ y_i=54$ and for $y_i= 15$
(i.e. $ \frac{\rho_{yi}}{\rho_{ri}}=8.3  \times10^{-20} $)
are explicitly shown in Figs. \ref{fig2}, \ref{fig3} and \ref{fig4}.
The dynamical evolution has a desired sequence of tracking solutions,
i.e., during the radiation era the  YMC
follows the radiation as $\rho_y\propto \rho_r \propto a(t)^{-4}$,
and then during the matter era it follows
the matter approximately as $\rho_y \propto \rho_m \propto a(t)^{-3}$,
and rather later around $z\sim 0.6$
it  becomes dominant.
Therefore, the matter era is long enough
for formation of large scale structure.
The fractional densities $\Omega_y=0.73$
and $\Omega_m=0.27$ are achieved at $z=0$ ($y=-0.97$)
as the current status of the Universe.
Notice that this tracking behavior is  always achieved
for any initial $\rho_{yi}$ in Eq.(\ref{yi})
whose range stretches over $\sim 25$ order in magnitude.
For a smaller initial $\rho_{yi}$, say $y_i\sim 15$,
$\rho_y(t)$ only tracks $\rho_r(t)$ for a shorter period
correspondingly, and then becomes the constant.
Moreover,
the  dynamical equation for $(y,x)$
at $t\rightarrow \infty$ has a fixed point $(y_f,x_f)=(-1.043, 0.114)$,
which has been examined to be stable, i.e.,
the solution is an attractor.
So, by the criteria mentioned in introduction,
the coincidence problem is solved also in this 2-loop model.
Furthermore, the EoS of YMC
crosses $w=-1$ around $z \simeq 1.7$ for the initial $y_i=54$,
but for smaller $y_i$ the crossing occurs earlier.
For any initial YMC, $w \simeq -1.14$ at $z=0$ is obtained,
which qualitatively agrees with
the fittings of preliminary observational data
\cite{riess06} \cite{alam}  \cite{Knop} \cite{Macorra}.
These behaviors are quite similar
to those of the 1-loop model \cite{zhang06},
even though the expressions involved in the 2-loop model
are much lengthier.

Parallel to the above of a constant $\gamma$,
the desired evolution is also realized
in an extended range of parameters,
$\gamma \simeq (0 \sim  0.6)$, and $\delta \simeq ( 3 \sim 7)$.
By calculation, we find  that  larger $\delta$ and $\gamma $
yield a lower $w$ at $z=0$.
For instances, at a fixed $\gamma =0.5$,
 $\delta =7\Rightarrow w\simeq -1.18$,
more interestingly, at a fixed $\delta =3$,
$\gamma=0.1\Rightarrow w\simeq -1.02$,
consistent with the $\Lambda$-model,
indicated by WMAP \cite{spergel}, SNLS \cite{Astier} and ESSENCE \cite{vasey},
and the crossing of $-1$ occurs rather late at $z\sim 0.35$,
quite close to the fitting of Gold+HST sample  \cite{riess06}.
The desired evolution can be also realized
for $\gamma$ being a generic function of the YMC $F$.
We have examined a variety of couplings,
say  $\gamma = 0.13 e^{y/4} (2+y)/(1+y)$
describing the YMC decaying into fermions
and  gauge bosons \cite{zhang-pairs},
and $\gamma=1.18e^y $ \cite{zhang06}, etc..
In the simple case of non-coupling, $\gamma=0$,
the dynamic evolution is similar to the coupling cases,
but   $w$ does not cross, only approaches to  $-1$.
Thus, if future observations do confirm EoS $w<-1$,
then the coupling  $\gamma>0$ is needed in our model;
otherwise, $\gamma=0$.

\begin{center}
{\em\Large 4. Summary}
\end{center}

The dark energy model based on the 2-loop quantum effective YMC,
under generic initial conditions and various forms of coupling,
has a stable tracking solution of dynamics,
thus naturally solves the coincidence problem.
Moreover, with coupling to the matter,
EoS of YMC  crosses  $-1$
and  $w\simeq -1.1 $ at $z=0$.
In this model the matter era ends rather late at $z\sim 0.6$,
so the matter era is long enough for
the structure formation.
Besides, since the YMC is subdominant during the early stages,
so the nuclear synthesis and the recombination
occur as in the standard Big Bang cosmology.
These conclusions are the same as for the 1-loop model.
With the Lagrangian being fixed by
quantum corrections,
the model depends on only two physical parameters:
the coupling $\Gamma\simeq (0\sim  0.6)H$ and
the energy scale $\kappa^{1/2}$.
As said earlier,
since the ``fine tuning''  problem is not solved by the model,
so the  energy scale
needs to be viewed as a new scale of physics.

ACKNOWLEDGMENT:  We thank the referee for helpful discussions.
Y.Zhang's work was supported by the CNSF No.10773009,
SRFDP, and CAS.

%\newpage

\newpage

\begin{figure}
%\centerline{\includegraphics[width=10cm]{fig1.eps}}
\caption{  \label{fig1}
$\rho_y$, $p_y$, and $w$ as functions of the variable $y=\ln(F/\kappa^2)$.
The  2-loop  model (solid) is compared with the 1-loop one (dotted)
\cite{zhang06}.}
\end{figure}
\begin{figure}
%\centerline{\includegraphics[width=10cm]{fig2.eps}}
\caption{ \label{fig2}
The evolution of energy densities
in the 2-loop  model with $\gamma =0.5$ and $\delta=3.0$.
Starting  from  $z_i=10^8$
the evolutions for two cases $y_i=15$ and $54$ are given.
For a whole wide range of initial YMC
$\rho_{yi}= (10^{-27}\sim \, 10^{-2})\rho_{ri}$,
there always exists a tracking solution,
and $\rho_y(t)$ becomes dominant around
$z\sim 0.6$. Due to the coupling, $\rho_m(t)$ will level off
in future. }
\end{figure}
\begin{figure}
%\centerline{\includegraphics[width=10cm]{fig3.eps}}
\caption{ \label{fig3}
The evolution of fractional energy densities
in the same model as Fig..\ref{fig2}.
At $z \simeq 3454$ the equality of radiation-matter $\Omega_m=\Omega_r$
occurred,
and the values
$\Omega_y=0.73$ and $\Omega_m=0.27$ are arrived at $z=0$.}
\end{figure}
\begin{figure}
%\centerline{ \includegraphics[width=10cm]{fig4.eps}}
\caption{  \label{fig4}
The evolution of EoS in the same model
as Fig.\ref{fig2} and Fig.\ref{fig3}.
In early stage at $z \rightarrow \infty$,
the YMC behaves like a radiation, $w \rightarrow1/3 $.
Later on $w$ reduces smoothly.
Due to the decay of YMC,
EoS crosses $-1$, and  $w\simeq -1.14$ at $z=0$.}
\end{figure}

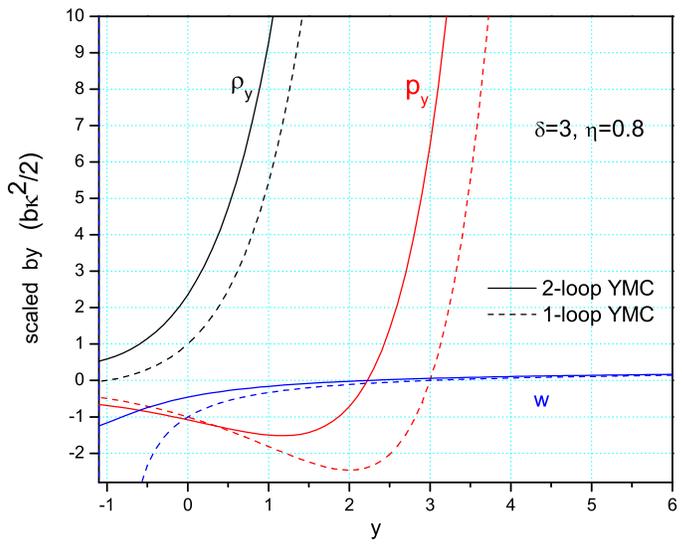
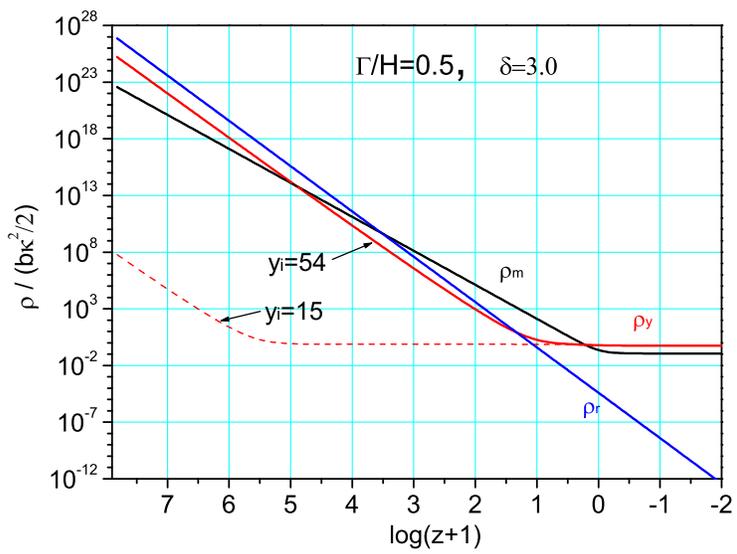
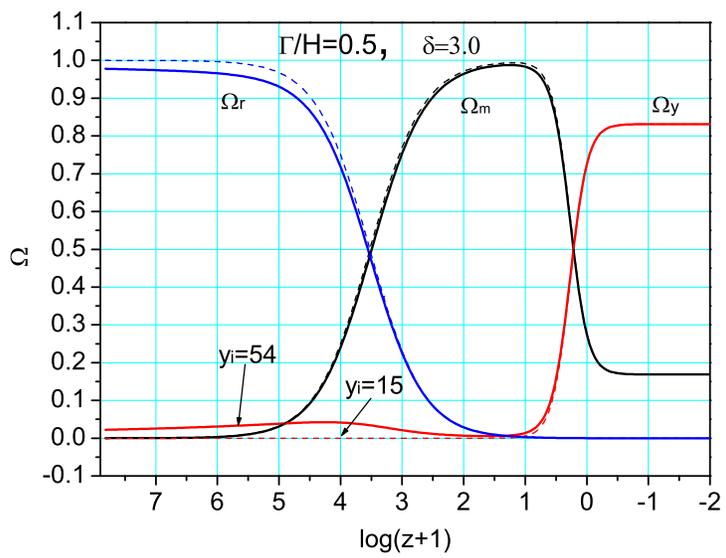
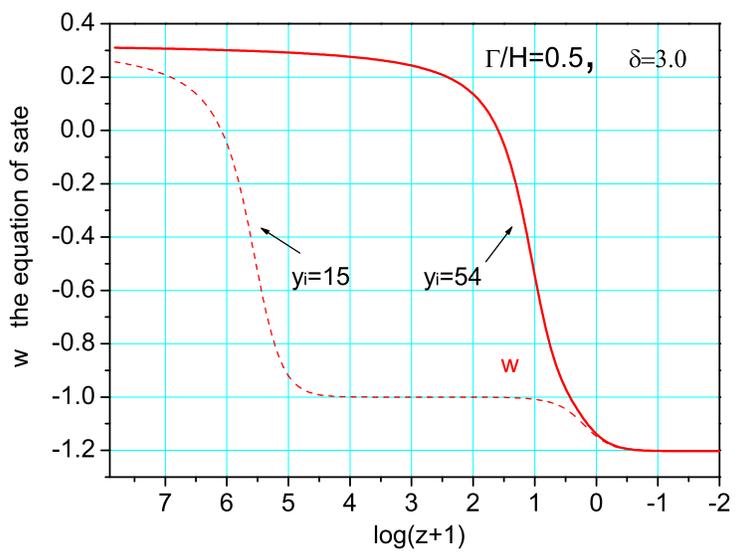
\end{document}